\documentclass[12pt]{article}
\usepackage{graphicx}
\usepackage{amssymb}

\usepackage{setspace}

\usepackage[dvips]{epsfig}
\usepackage{subfigure}

\begin{document}


\title{Two-strain ecoepidemic systems: the obligated mutualism case
}

\author{Chiara Bosica, Alessandra De Rossi, Noemi Lucia Fatibene,\\
Matteo Sciarra, Ezio Venturino\thanks{Corresponding author.
Email: ezio.venturino@unito.it}\\
Dipartimento di Matematica ``Giuseppe Peano'',\\
Universit\`{a} di Torino, Italy.
}
\date{}
\maketitle

{\textbf{Abstract}}
We present a model for obligated mutualistic associations, in which two transmissible diseases are allowed to infect just one
population. As the general model proves too hard to be fully analytically investigated, some special cases are analysed.
Among our findings, the
coexistence of the two strains does not appear possible, under the model assumptions. Furthermore, in
particularly unfavorable circumstances the ecosystem may disappear.
In this respect, an accurate computation of the basin of attraction of the origin is provided
using novel techniques.
For this obligated mutualistic system the presence of the
diseases appears to be less relevant than in many other circumstances in ecoepidemiology,
including also the case of facultative symbiotic associations.



\section{Introduction}

Symbiotic associations occur frequently in nature, although in population theory mathematical models focus generally more on
competing situations or predator-prey interactions:
the classical examples are the anemone-damselfish and the ant-plant interactions leading to pollination, \cite{B}.
In the latter context, for instance moths (of genus Tegeticula) pollinate yuccas, \cite{A}.
Other known associations involve mycorrizhal fungi,
fungus-gardening ants, mixed feeding flocks of birds dispersing seeds of {\it{Casearia corymbosa}}
in Costa Rica, \cite{J80}.
Commensalism and symbiotic populations have been considered within food chains where some
of the other populations are in competition with each other,
\cite{DDD,KF02,KF89,Z,GL,GYW,RFA}.
A recent contribution along these lines is \cite{C_et_al}, considering symbiotic models at various trophic levels in food chains.

Ecoepidemiology is a rather recent field of study, investigating the effect that epidemics have on the
underlying demographic populations interactions.
Many papers by now have been devoted to the study of
ecoepidemic systems based on predator-prey or competing demographics.
For an account of some of the early developments in this field,
see Chapter 7 of \cite{MPV}.
In fact, diseases cannot be ignored in ecosystems. A whole wealth of possible ailments affecting populations in
acquatic, terrestrial or avian environments is contained in \cite{Gull}.

Specific examples involving populations living in symbiosis can also be found,
e.g. several mushrooms ({\it{Cantharellus cibarius}},
{\it{Boletus}} spp., {\it{Amanita}} spp.) with chestnut trees ({\it{Castanea sativa}}).
The disease in this case is represented by chestnut cancer ({\it{Endothia parasitica}}).
Symbiotic associations are common among bacteria alone, \cite{ZL}, bacteria together with other organisms
\cite{B-EG,SC}, plants \cite{Pa} and plants and mushrooms \cite{MK} and these symbiotic systems
may affect the whole ecosystem in which they thrive, \cite{FCMCR}.
Other instances are
the soil nematode {\it{Caenorhabditis elegans}}, that transfers the rhizobium species
{\it{Sinorhizobium melotiti}} to the roots of the legume {\it{Medicago truncatula}} \cite{AFKSSK,HP},
the L-form bacteria in non-pathogenic symbiosis with several
plants that allow the latter to resist other bacterial pathogens \cite{WF}.
These considerations were the underlying motivations for studying a symbiotic situation encompassing diseases, \cite{EV:07}.
The investigation has been extended in \cite{NOVA}, assuming a Holling type II term for the possible mutual rewards of the symbiotic populations.

Other developments in epidemiology have dealt with the case of two pathogens affecting together a host.
Ecoepidemic situations of this type that have been investigate previously consist of two diseases that are assumed
to spread in a predator-prey community, affecting the predators, \cite{Roman}, or the prey, \cite{Elena1,Elena2}.
In this paper, our aim is to further extend the work \cite{NOVA}, by considering different strains that affect the symbiotic environment.
Specifically, we consider a general symbiotic model and some particular cases. The fundamental assumption that relates the two diseases
is that they do not interfere with each other. This means that they cannot both affect at the same time the same individual, i.e. there is no coinfection,
nor superinfection, i.e. once an individual gets a disease, he is prevented from
being affected also by the other one. The mutualistic association is obligated, i.e. without the other one, each population would not thrive alone.
Possible extensions that are not discussed here are represented instead by facultative mutualism;
also the assumption that diseased individuals do not receive benefits from interactions with the other population could be removed.

The paper is organized as follows. We briefly discuss the demographic model as a further reference, then introduce in Section 3 the general model.
As a complete analysis is not possible, in Sections 4 and 5 we investigate two particular cases, restricting somewhat the infected from taking part
in the association. A final discussion concludes the paper.

\section{Preliminaries}

\begin{figure}[htbp]
\centering
\includegraphics[width=9cm]{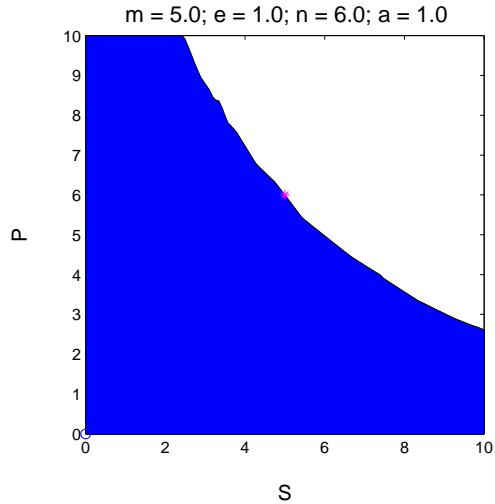}
\caption{The blue colored region represents the basin of attraction of the origin in the $SP$ phase plane.}
\label{fig:separ2}
\end{figure}

Before introducing the ecoepidemic models, for comparison purposes, we briefly discuss their underlying demographic model,
i.e. the model without the infected individuals,
\begin{eqnarray}\label{Mod_3}
\frac{dS}{dt}=-nS+aSP,\quad \frac{dP}{dt}=-mP+eSP.
\end{eqnarray}
All parameters here and in the next Sections are always assumed to be nonnegative.
System (\ref{Mod_3}) has only two equilibria, the origin $\widehat E_0=(0,0)$ and the coexistence point $\widehat E_1=(me^{-1},na^{-1})$.
It is very simple to write down its Jacobian $J_d$,
$$
J_d=
\left[
\begin{array}{cc}
-n+aP & aS\\
Pe & -m+eS 
\end{array}
\right] 
$$
and from its evaluation at $\widehat E_0$ to find the eigenvalues $-n$ and $-m$, while the evaluation at $\widehat E_1$ gives the eigenvalues $\pm\sqrt{mn}$.
It follows that the coexistence equilibrium is unstable, namely a saddle, and the origin is always stable. Thus the phase plane is partitioned into
two domains, one for which the origin is an attractor, and the other one in which the trajectories ultimately drift to infinity. Therefore to prevent the system's extinction, 
in practical situations it is important to assess the basin of attraction of the origin.
To this end,
based on the very recent algorithm
presented in \cite{ABCDe,ICNAAM}, we show in Figure \ref{fig:separ2} the separatrix of these domains, obtained for the following set of parameter
values $m = 5.0$, $e = 1.0$, $n = 6.0$, $a = 1.0$.


The result is partly unsatisfactory, since from the biological point of view the ecosystem is bound to disappear, if the population levels are low,
or better said if they fall in the domain of attraction of the origin, otherwise they explode. From the ecological point of view this latter phenomenon is
impossible, since finite resources cannot sustain an ever increasing population. However, it is imbedded in the assumptions of the model, which are
kept at a minimal number, in order to better analyse the ecoepidemic models that follow and compare their results with those of the
underlying purely demographic model. In this way the disease influence on the environment transpires more clearly.

\section{The general model}

As before, let $S$ and $P$ be the two symbiotic populations. We assume that two recoverable diseases spread by contact among the $S$ population, giving rise to
infected individuals of type $H$ and $Y$.
We assume that there is neither coinfection nor superinfection, i.e. whenever one individual is infected by one strain, it cannot catch the other
disease and become infected with both, nor can he get the second disease and the latter replace the first one.
The model, in which all parameters are assumed to be nonnegative, reads
\begin{eqnarray}\label{Mod_general}
& & \frac{dS}{dt}=-nS-\lambda HS-\beta YS+aSP+\xi H+\varphi Y,\\ \nonumber
& &\frac{dH}{dt}=\lambda HS-\mu H-\xi H+qHP,\\ \nonumber
& &\frac{dY}{dt}=\beta YS-\nu Y-\varphi Y+rYP,\\ \nonumber
& &\frac{dP}{dt}=-mP+eSP+fHP+gYP.
\end{eqnarray}

The first equation models the dynamics of the $S$ population. It dies out at an exponential rate $n$, and reproduces only in presence of the other
mutualistic population $P$, at rate $a$. By causal encounters with infected individuals of type $H$ and $Y$, a susceptible can then become infected,
at rates $\lambda$ and $\beta$ respectively. Finally, individuals of the latter two classes that recover from the disease reenter the susceptible
class.

The second equation considers the infected $H$; they are recruited among the susceptibles at rate $\lambda$, as mentioned, they recover at rate $\xi$
and are subject to natural plus disease-related mortality $\mu$.
The third equation contains a similar dynamics for the infected $Y$. Note that in this case the total mortality rate is named $\nu$ and the recovery
rate is expressed by the parameter $\varphi$.

Finally, the fourth equation shows the behavior of the mutualistic population $P$, which in absence of individuals of the mutualistic population
dies out at an exponential rate $m$, and reproduces, when the symbiotic population is present, at rate $e$.

The system's equilibria are the following points. The origin $E_0=(0,0,0,0)$, and the coexistence equilibrium $E_4=\left(S_4,H_4,Y_4,P_4\right)$
with population values
$$S_4=\frac{\nu q+\varphi q-r\mu-r\xi}{\beta q-r\lambda}, \quad P_4=\frac{\mu\beta+\xi\beta-\lambda\nu-\lambda\varphi}{\beta q-r\lambda},$$
$$H_4=\frac{(ng+\beta m+\varphi e)S_4-\beta eS_4^2-agP_4S_4-\varphi m}{(\beta f-\lambda g)S_4+\xi g-\varphi f},$$
$$Y_4=\frac{e\lambda S_4^2-(m\lambda+e\xi+fn)S_4+afS_4P_4+m\xi}{(\beta f-\lambda g)S_4+\xi g-\varphi f}. $$
and then $E_1=\left( me^{-1},0,0, na^{-1} \right)$,
$$E_{2\pm}=\left(z_{2,\pm},\frac{1}{f}(m-ez_{2,\pm}),0,\frac{1}{q}(\mu+\xi-\lambda z_{2,\pm})\right),$$
$$E_{3\pm}=\left(z_{3,\pm},0,\frac{1}{g}(m-ez_{3,\pm}),\frac{1}{r}(\nu+\varphi-\beta z_{3,\pm})\right),$$
where $z_{2,\pm}$ are the roots of
\begin{equation}\label{q2}
R_2(Z)\equiv (a\lambda f-\lambda qe)Z^2+(nqf+\lambda qm+\xi qe-a\mu f-a\xi f)Z-\xi qm =0.
\end{equation}
while $z_{3,\pm}$ solve
\begin{equation}\label{q1}
R_1(Z)\equiv(a\beta g-\beta re)Z^2+(nrg+\beta rm+\varphi re-a\nu g-a\varphi g)Z-\varphi rm =0,
\end{equation}

$E_0$ are $E_1$ always feasible.

To investigate feasibility of $E_{2+}$, let $\Delta_{2}$ be the discriminant of (\ref{q2}). Imposing
$z_{2,+}>0$ we have $nqf+ \lambda qm-a \mu f-\sqrt{\Delta_{2}}< \xi (af-eq)$, which together with the nonnegativity conditions
for the remaining populations, that give $e< m z_{2,+}^{-1}$ and $ \lambda z_{2,+}- \xi< \mu$, provides bounds for $\xi$. Namely, for 
\begin{eqnarray}\label{E6+_feas2a}
f>\max \left\{ \frac{\mu eq-\lambda mq+\sqrt{\Delta_{2}}}{nq},\frac{\mu e^2q+e\sqrt{\Delta_{2}}}{m \lambda a+nqe} \right\}
\end{eqnarray}
we find
\begin{eqnarray}\label{E6+_feas2b}
\frac{nqf+\lambda qm- a \mu f -\sqrt{\Delta_{2}}}{af-eq}< \xi < 
\frac{nqf-m \lambda q-a \mu f-\sqrt{\Delta_{2}}}{af-eq}+\frac{2m \lambda af}{e(af-eq)}, 
\end{eqnarray}
which to be consistent, requires $af>eq$.

In case instead that
\begin{equation}\label{E6+_feas3b}
\frac{\mu e^2q+e\sqrt{\Delta_{2}}}{m \lambda a+nqe}<f \leq \frac{\mu eq- \lambda mq+\sqrt{\Delta_{2}}}{nq},
\end{equation}
we find the interval
\begin{eqnarray}\label{E6+_feas3a}
\frac{2 \mu eq-a \mu f-nqf- \lambda qm+\sqrt{\Delta_{2}}}{af-eq}< \xi <
\frac{nqf-m \lambda q-a \mu f-\sqrt{\Delta_{2}}}{af-eq}+\frac{2m \lambda af}{e(af-eq)}, 
\end{eqnarray}
which implies
$$
f>\frac{\mu e^2q+e\sqrt{\Delta_{2}}}{m \lambda a+nqe},
$$
if we take $af>eq$, or a contradiction with (\ref{E6+_feas3b}) in the opposite case.
In summary $E_{2+}$ is feasible if
\begin{equation}\label{E6+_feas1}
af>eq
\end{equation}
together with either (\ref{E6+_feas2a}) and (\ref{E6+_feas2b}), or together with (\ref{E6+_feas3a}) and (\ref{E6+_feas3b}).
The inequality (\ref{E6+_feas3b}) to be true implies $\mu eaq-m\lambda aq-nq^2e+a\sqrt{\Delta_2}>0$.

For $E_{2-}$ we have similar results. We need again (\ref{E6+_feas1}) together with the same two alternative sets of conditions
(\ref{E6+_feas2a}) and (\ref{E6+_feas2b}), or respectively (\ref{E6+_feas3a}) and (\ref{E6+_feas3b}), in which however the plus sign in
front of the square root is replaced by a minus.

Note that when these two points, $E_{2\pm}$ coalesce, the same feasibility conditions still hold, in a simplified form: it is enough to set
$\Delta_2=0$ in all the previous formulae.

For the pair of equilibria $E_{3\pm}$ similar steps lead to the following feasibility conclusions. If we denote by
$\Delta_3$ the discriminant of (\ref{q1}), for $E_{3-}$ we need
\begin{equation}\label{E3+_feas1}
ag>er
\end{equation}
and
\begin{eqnarray}\label{E3+_feas2a}
g>\max\left\{\frac{\nu er-\beta mr-\sqrt{\Delta_3}}{rn},\frac{\nu e^2r-e\sqrt{\Delta_3}}{m\beta a+nre}\right\}
\end{eqnarray}
together with
\begin{eqnarray}\label{E3+_feas2b}
\frac{nrg+\beta rm-a\nu g+\sqrt{\Delta_3}}{ag-er}<\varphi <\frac{2m\beta ag -m\beta er +enrg-ea\nu g+e\sqrt{\Delta_3}}{e(ag-er)}
\end{eqnarray}
or (\ref{E3+_feas1}) together with
\begin{eqnarray}\label{E3+_feas3a}
\frac{\nu e^2r-e\sqrt{\Delta_3}}{m\beta a+nre}<g\leq \frac{\nu er-\beta mr-\sqrt{\Delta_3}}{nr},
%
\end{eqnarray}
which requires $\nu era-m\beta ar-nr^2e-a\sqrt{\Delta_3}>0$, and
\begin{eqnarray}\label{E3+_feas3b}
\frac{\nu (2er-ag) -r(ng+\beta m) -\sqrt{\Delta_3}}{ag-er}<\varphi <\frac{m\beta (2ag - er) +eg (nr-a\nu) +e\sqrt{\Delta_3}}{e(ag-er)}.
\end{eqnarray}
For $E_{3+}$ again to obtain feasibility conditions it is enough to change the signs of the square root terms, or to set it to zero in case
the two points coalesce.

To assess stability analytically is too complex. From our simulations, it seems that all these points are unstable, the system tending either
to the origin, or eventually the trajectories becoming unbounded.

To better analyse the system, we now turn to the analysis of some simplified cases.

\section{No intermingling with infected allowed}

We make here the simplifying assumption that the second population $P$ does not interact
with the infected both strains of the population $S$, because they can be recognized and therefore avoided, for instance.
This corresponds to setting $f=q=g=r=0$ in (\ref{Mod_general}). Explicitly, we thus have
\begin{eqnarray}\label{Mod_1}
 & & \frac{dS}{dt}=-nS-\lambda HS-\beta YS+aSP+\xi H+\varphi Y,\\ \nonumber
 & & \frac{dH}{dt}=\lambda HS-\mu H-\xi H,\\ \nonumber
 & & \frac{dY}{dt}=\beta YS-\nu Y-\varphi Y,\\ \nonumber
 & & \frac{dP}{dt}=-mP+eSP.
\end{eqnarray}

In this case, however, there are only the two feasible equilibria
$Q_0\equiv E_0=(0,0,0,0)$ and $Q_1\equiv E_1 =\left( me^{-1},0,0,na^{-1} \right)$,
that are always feasible. They clearly coincide with those of (\ref{Mod_3}), except that have two dimensions, i.e. two populations, more, the
infected, although the latter are at zero level. We note thus that either one of nor both the two diseases cannot survive in this system,
they are eradicated. From the epidemiological point of view this is a very important result, subject of course to the rather peculiar assumptions
of the underlying demographic model, i.e. Malthusian growth, or in terms of ecosystem, of the fact that that this is an obligated mutualism.
The mathematical reason for which equilibria with diseases at positive level are
not sustainable, is that the points $Q_2^*$ with only nonzero populations $S_2^*=(\nu+\varphi)\beta^{-1}$, $Y_2=-n\nu^{-1}S_2^*$
and $Q_3^*$ with the nonzero population levels $S_3^*=(\mu+\xi)\lambda^{-1}$, $H_3^*=-n\mu^{-1}S_3^*$ have both a negative component.

The Jacobian of (\ref{Mod_1}) is
$$
J=
\left[
\begin{array}{cccc}
-n-\lambda H-\beta Y+aP & - \lambda S +\xi & -\beta S+\varphi & aS \\
\lambda H & \lambda S -\mu -\xi  & 0 & 0 \\
\beta Y & 0 & \beta S- \nu - \varphi & 0 \\
Pe & 0 & 0 & -m+eS 
\end{array}
\right] .
$$
Its evaluation at $Q_0$ gives the eigenvalues $- \nu - \varphi$, $-n$, $- \mu - \xi$, $-m$, which are all negative. Thus the origin is once again
unconditionally stable. Evaluation at $Q_1$ leads instead to the eigenvalues
$\pm\sqrt{mn}$, $(\beta m - \nu e -\varphi e)e^{-1}$, $(\lambda m - \mu e - e \xi)e^{-1}$.
Since the first two are those inherited from the corresponding equilibrium $\widehat E_1$, and one of them is positive, we conclude that also
$Q_1$ is unstable.
Thus the disease in this context does not really change the system's behavior.

\begin{figure}[htbp]
\centering
\includegraphics[width=9cm]{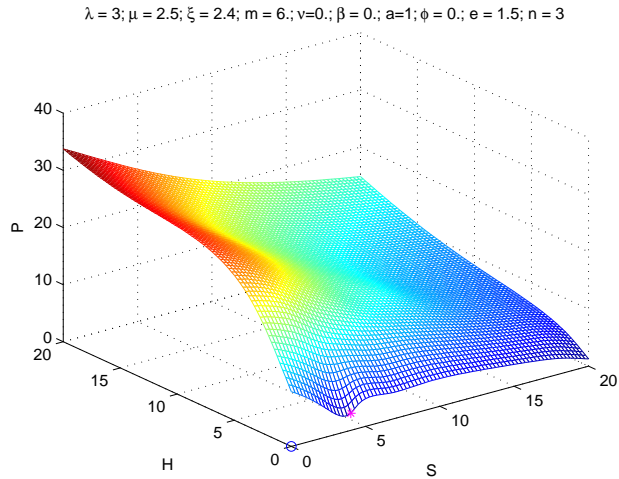}
\caption{The region below the surface represents the basin of attraction in the $SHP$ phase subspace $Y=0$. The star represents the
projection of the $Q_1\equiv E_1=(4,0,0,3)$ saddle point.}
\label{fig:separ3}
\end{figure}

Due to the threat of a vanishing ecosystem, a small region of attraction of the origin is desirable.
Reliable procedures for its determination have been devised, by computation of the separatrix surface, \cite{Emma}.
In Figure \ref{fig:separ3} we show the picture of the basin of attraction of the origin, which lies
below the surface,
for the hypothetical parameter
values $\lambda = 3$, $\mu = 2.5$, $\xi = 2.4$, $m = 6.$, $\nu=0.$, $ \beta = 0.$, $a=1$, $\phi = 0.$, $e = 1.5$, $n = 3$.

\section{Infected do not get reward from symbiosis}
In this case the $P$ population gains from the interactions also with the infected individuals of the first one,
but the latter do not feel the benefit of the mutualism. In other words, we set only $r=q=0$ in (\ref{Mod_general}).
Thus, explicitly,
\begin{eqnarray}\label{Mod_2}
& & \frac{dS}{dt}=-nS-\lambda HS-\beta YS+aSP+\xi H+\varphi Y,\\ \nonumber
& &\frac{dH}{dt}=\lambda HS-\mu H-\xi H,\\ \nonumber
& &\frac{dY}{dt}=\beta YS-\nu Y-\varphi Y,\\ \nonumber
& &\frac{dP}{dt}=-mP+eSP+fHP+gYP.
\end{eqnarray}
The equilibria are now
\begin{eqnarray*}
\widetilde E_0=(0,0,0,0),\ \ \widetilde E_1=\left( \frac{m}{e},0,0, \frac{n}{a} \right),\\
\widetilde E_2=\left(\frac{\mu+\xi}{\lambda}, \frac{m \lambda -e \mu -e \xi }{f\lambda},0,
\frac{m \lambda \mu -e \mu^2-e \xi \mu +fn \mu +fn \xi }{fa(\mu +\xi)}\right),\\
\widetilde E_3=\left(\frac{\nu +\varphi}{\beta},0,\frac{m \beta -e \nu -e \varphi }{g\beta},
\frac{m\beta\nu -e \nu^2-e\varphi\nu+gn\nu+gn\varphi}{ga(\nu+\varphi)}\right).
\end{eqnarray*}
Again, note that two more disease-unaffected-population-free equilibria
with endemic disease are not feasible, and coincide with those of the former model (\ref{Mod_1}),
$E_2^*\equiv Q_2^*$ and $E_3^*\equiv Q_3^*$.

While the origin and $\widetilde E_1$ are always feasible, the remaining points are only if some conditions hold. Specifically,
feasibility conditions can now be explicitly stated as follows:
for $\widetilde E_2$ we have
\begin{equation}\label{E3_feas}
e\leq \frac{m\lambda}{\mu + \xi},\quad n\geq \frac{\mu (e\mu -m\lambda +e\xi)}{f(\mu + \xi)},
\end{equation}
while for $\widetilde E_3$ we find
\begin{equation}\label{E5_feas}
e\leq \frac{m\beta}{\nu + \varphi},\quad n\geq \frac{\nu (e\nu -m\beta +e\varphi)}{g(\nu + \varphi)}.
\end{equation}

The system's (\ref{Mod_2}) Jacobian is
$$
J=
\left[
\begin{array}{cccc}
J_{11} & - \lambda S +\xi & -\beta S+\varphi & aS \\
\lambda H & \lambda S -\mu -\xi  & 0 & 0 \\
\beta Y & 0 & \beta S- \nu - \varphi & 0 \\
Pe & Pf & Pg & J_{44}
\end{array}
\right] ,
$$
with
$J_{11}=-n-\lambda H-\beta Y+aP$, $J_{44}=-m+eS+fH+gY$.

It is easy to verify that $E_0$ and $E_1$ retain the stability properties respectively of $Q_0$ and $\widehat E_0$, as well as $Q_1$ and $\widehat E_1$.
The eigenvalues of the Jacobian evaluated at $\widetilde E_2$ are
$$\frac{\beta\mu+\beta\xi - \nu\lambda - \varphi\lambda}{\lambda}$$
and the roots of the cubic monic polynomial ($c_3=1$),
$p(t)=\sum_{i=0}^{3}c_it^i$, with
\begin{eqnarray*}
c_0 & = & \frac{1}{f\lambda}(e\mu^2fn-e^2\mu^3-2e^2\mu^2\xi +2m\lambda e\mu^2+2e\mu fn\xi-m^2\lambda^2\mu-m\lambda fn\mu\\
 && 2m\lambda e\xi\mu-e^2\xi^2\mu +e\xi^2 fn+m\lambda fn\xi),\\
c_1 &=& \frac{\mu^2e^2-\lambda e\mu^2+\mu m\lambda^2-\lambda e \xi\mu -\mu m\lambda e-\mu efn+\mu e^2\xi-efn\xi}{f\lambda},\\
c_2&=&\frac{\xi(m\lambda-e\mu-e\xi)}{f(\mu+\xi)}.
\end{eqnarray*}
The Routh-Hurwitz stability criterion requires strict positivity for the following quantities:
$$
D_{2,1}=\left|c_2\right|,\quad
D_{2,2}=\left|\begin{array}{cc}c_2 & c_0\\c_3 & c_1\end{array}\right|,\quad
D_{2,3}= \left|\begin{array}{ccc}c_2 & c_0 & 0\\c_3 & c_1 & 0 \\ 0 & c_2 & c_0 \end{array}\right|.
$$

Now $D_{2,1}>0$ is a consequence of the strict feasibility condition (\ref{E3_feas}).

For $D_{2,2}>0$, we have
\begin{eqnarray*}
D_{2,2} = \frac{1}{\lambda(\mu+\xi)f^2}(m\lambda-e\mu-e\xi)(e^2\mu^2\xi+e^2\xi^2\mu-f\mu^3e-\lambda e\mu^2\xi \\
-2ef\mu^2\xi-fe\mu\xi^2-fne\mu\xi-e\lambda\mu\xi^2-me\lambda\mu\xi-fen\xi^2\\
+nf^2\mu^2+mf\lambda\mu^2+m\lambda^2\mu\xi+mf\lambda\mu\xi+2nf^2\mu\xi+f^2n\xi^2).
\end{eqnarray*}
The denominator is positive as well as the first factor, when the first (\ref{E3_feas}) is
satisfied as a strict inequality. The second factor is a quadratic in the parameter $e$,
whose roots we indicate by $e_1$ and $e_2$. The second factor is positive for the values: $e<e_2$ or $e>e_1$ in the case of real roots,
($e_2 < e_1$); for every $e\neq e_1$ in case of a double real root; for every value of $e$ when the roots are complex.

Finally, to have $D_{2,3}>0$ we need
\begin{eqnarray*}
D_{2,3} = \frac{1}{\lambda^2(\mu+\xi)f^3}(m\lambda-e\mu-e\xi)^2(e^2\mu^2\xi+e^2\xi^2\mu-f\mu^3e-\lambda e\mu^2\xi \\
-2ef\mu^2\xi-fe\mu\xi^2-fne\mu\xi-e\lambda\mu\xi^2-me\lambda\mu\xi-fen\xi^2\\
+nf^2\mu^2+mf\lambda\mu^2+m\lambda^2\mu\xi+mf\lambda\mu\xi+2nf^2\mu\xi+f^2n\xi^2)\\
\times (e\mu^2-m\lambda\mu+e\xi\mu-fn\mu-fn\xi) >0.
\end{eqnarray*}
Here too the denominator is positive, the first factor is when
$$
e\neq  \frac{m\lambda}{\mu + \xi},
$$
the second one coincides with the second one of $D_{2,2}$, the third one is positive for
$$
e>\frac{m\lambda\mu+fn\mu+fn\xi}{\mu(\mu+\xi)}.
$$

By combining all these cases we find $D_{2,3}>0$ for the following different cases
\begin{enumerate}
\item for different real roots of the second factor, we need one of the following two alternative situations
\begin{eqnarray*}
e>\frac{m\lambda\mu+fn\mu+fn\xi}{\mu(\mu+\xi)}, \quad e\neq  \frac{m\lambda}{\mu + \xi}, \quad e<e_2;\\
e>e_1, \quad e>\frac{m\lambda\mu+fn\mu+fn\xi}{\mu(\mu+\xi)}, \quad e\neq  \frac{m\lambda}{\mu + \xi},
\end{eqnarray*}
or all the following conditions
$$
e\neq  \frac{m\lambda}{\mu + \xi}, \quad e_2<e<e_1, \quad e<\frac{m\lambda\mu+fn\mu+fn\xi}{\mu(\mu+\xi)};
$$
\item when there are two double roots, we need all the following conditions
$$
e\neq  \frac{m\lambda}{\mu + \xi}, \quad e\neq e_1\equiv e_2, \quad e>\frac{m\lambda\mu+fn\mu+fn\xi}{\mu(\mu+\xi)};
$$ 
\item for complex roots instead the required conditions are
$$
e\neq  \frac{m\lambda}{\mu + \xi}, \quad e>\frac{m\lambda\mu+fn\mu+fn\xi}{\mu(\mu+\xi)}.
$$
\end{enumerate}

But none of these conditions can hold, namely:
\begin{itemize}
\item the conditions $e_2<e<e_1$ cannot hold in view of the request $D_{2,2}>0$;
\item the remaining three cases require 
$$
e>\frac{m\lambda\mu+fn\mu+fn\xi}{\mu(\mu+\xi)}=\frac{m\lambda\mu}{\mu(\mu+\xi)}+\frac{fn(\mu+\xi)}{\mu(\mu+\xi)}
=\frac{m\lambda}{\mu+\xi}+\frac{fn}{\mu}>\frac{m\lambda}{\mu+\xi},
$$
which contradicts the feasibility condition (\ref{E3_feas}).
\end{itemize}
In conclusion, $D_{2,3}$ cannot be positive, so that $\widetilde E_2$ is unconditionally unstable for all parameter choices.

Note that Hopf bifurcations also cannot arise.
We would need $c_1c_2-c_0=0$, but solving it in terms of the parameter $n$, the value it must have so that purely imaginary eigenvalues arise is
$$
n=\frac{(m\lambda-e\mu-e\xi)(e\xi-f\mu-\lambda\xi-f\xi)\mu}{(\mu+\xi)(f\mu+f\xi-e\xi)f}.
$$
However, the feasibility condition (\ref{E3_feas}) for $\widetilde E_2$ in terms of $n$ requires
$$
n\geq \frac{\mu (e\mu -m\lambda +e\xi)}{f(\mu + \xi)}.
$$
Combining the two above conditions, we find
$$
\frac{e\xi-f\mu-\lambda\xi-f\xi}{e\xi-f\mu-f\xi}\geq 1,
$$
which implies
$e\xi-f\mu-\lambda\xi-f\xi\geq e\xi-f\mu-f\xi$ and finally $-\lambda\xi\geq 0$, which is impossible.

For $\widetilde E_3$ the situation is similar, we have one eigenvalue as follows
$$
\frac{\varphi \lambda -\beta \mu - \beta \xi + \nu \lambda }{\beta}
$$
and the roots of the cubic 
$q(t)= \sum_{i=0}^{3} h_it^i$ with coefficients $h_3=1$ and
\begin{eqnarray*}
h_0 &=& \frac{1}{g \beta}(2m \beta e \nu^2 -e^2 \nu^3 -2 e^2 \nu^2 \varphi +e \nu^2 gn +2e \nu g n \varphi -e^2 \varphi^2 \nu -m^2 \beta^2 \nu\\
&&+2m \beta e \varphi \nu -m \beta g n \nu -m \beta g n \varphi  +e \varphi^2 gn),\\
h_1&=&\frac{\nu^2e^2- \beta e \nu^2 +\nu m \beta^2 -\beta e \varphi \nu + \nu e^2 \varphi - \nu m \beta e- \nu egn -egn \varphi}{g \beta}, \\
h_2&=& \frac{\varphi (m \beta - e \nu - e \varphi)}{g( \nu + \varphi)}.
\end{eqnarray*}
Once again, $D_{3,1}=c_2>0$ follows from the strict first feasibility condition (\ref{E5_feas}).
We then need the positivity of
$$
D_{3,2}=
\left|
\begin{array}{cc}
c_2 & c_0 \\ c_3 & c_1
\end{array}
\right|
$$
which gives
\begin{eqnarray*}
D_{3,2} = \frac{1}{g^2(\nu+\varphi)\beta}(m \beta - e \nu - e \varphi)(e^2\nu^2\varphi+e^2\varphi^2\nu-e\nu^3g-2e\nu^2g\varphi \\
-e\nu^2\beta\varphi-e\nu g\varphi^2-e\varphi^2\nu\beta-m\beta e\varphi\nu-e\nu gn\varphi-e\varphi^2gn+m\beta g\nu^2\\
+\nu^2g^2n+\varphi m\beta\nu g+m\beta^2\nu\varphi+2\nu g^2n\varphi+g^2n \varphi^2)>0.
\end{eqnarray*}
The denominator is positive, the first factor is also when the first feasibility condition (\ref{E5_feas}) is strictly satisfied.
The second factor is a quadratic in $e$, whose roots are denoted $e_3$ and $e_4$. It is positive for $e<e_4$ o $e>e_3$ when $e_3 > e_4$ are the two real roots.
When they coincide, we need $e\neq e_3 \equiv e_4$; finally, for complex roots, $D_{3,2} >0$ unconditionally.

The third Routh-Hurwitz condition requires 
\begin{eqnarray*}
D_{3,3} = \frac{1}{g^3(\nu+\varphi)\beta^2}(m \beta - e \nu - e \varphi)^2 
(e\nu^2-m\beta\nu+e\varphi\nu-gn\nu-gn\varphi)\\
\times (e^2\nu^2\varphi+e^2\varphi^2\nu-e\nu^3g-2e\nu^2g\varphi-e\nu^2\beta\varphi-e\nu g\varphi^2-e\varphi^2\nu\beta\\
-m\beta e\varphi\nu-e\nu gn\varphi-e\varphi^2gn+m\beta g\nu^2+\nu^2g^2n+\varphi m\beta\nu g\\
+m\beta^2\nu\varphi+2\nu g^2n\varphi+g^2n \varphi^2)>0.
\end{eqnarray*}
If we take
$$
e\neq \frac{m\beta}{\nu + \varphi}
$$
the sign of $D_{3,3} $ depends only on the last two factors, and the last one is the second factor
of $D_{3,2} $ so that we find once again the roots $e_3$ ed $e_4$.
The third factor is positive whenever
$$
e> \frac{m \beta \nu + gn \nu +gn \varphi}{\nu ( \nu + \varphi)}.
$$

In summary,
$D_{3,3}>0$ in the following situations
\begin{enumerate}
\item for two real distinct roots $e_3$ and $e_4$ we either need both the following conditions
\begin{eqnarray*}
e> \frac{m \beta \nu + gn \nu +gn \varphi}{\nu ( \nu + \varphi)}, \quad e\neq \frac{m\beta}{\nu + \varphi}, \quad e<e_2;\\
e>e_1, \quad e> \frac{m \beta \nu + gn \nu +gn \varphi}{\nu ( \nu + \varphi)}, \quad e\neq \frac{m\beta}{\nu + \varphi},
\end{eqnarray*}
or, alternatively, we can also have 
$$
e\neq \frac{m\beta}{\nu + \varphi}, \quad e_2<e<e_1 \quad e< \frac{m \beta \nu + gn \nu +gn \varphi}{\nu ( \nu + \varphi)};
$$
\item for real identical roots, $e_3\equiv e_4$ we need
$$
e\neq \frac{m\beta}{\nu + \varphi}, \quad e\neq e_1 \equiv e_2, \quad e> \frac{m \beta \nu + gn \nu +gn \varphi}{\nu ( \nu + \varphi)};
$$
\item for complex roots, instead the requirement is
$$
e\neq \frac{m\beta}{\nu + \varphi}, \quad
e> \frac{m \beta \nu + gn \nu +gn \varphi}{\nu ( \nu + \varphi)}.
$$
\end{enumerate}

Again, none of the above conditions can hold. In fact,
\begin{itemize}
\item whenever $e_2<e<e_1$, we contradict the statement that $D_{3,2}>0$; 
\item in the remaining cases we want
$$
e> \frac{m \beta \nu + gn \nu +gn \varphi}{\nu ( \nu + \varphi)}=\frac{m \beta \nu}{\nu ( \nu + \varphi)} + \frac{gn (\nu + \varphi)}{\nu (\nu + \varphi)}
= \frac{m \beta}{( \nu + \varphi)} + \frac{gn}{\nu}> \frac{m \beta}{( \nu + \varphi)}
$$
which clashes with the first feasibility condition (\ref{E5_feas}) for $\widetilde E_3$.
\end{itemize}

In summary, $\widetilde E_3$ is always unstable.
Here too, Hopf bifurcations are impossible, since $h_1h_2-h_0=0$ in term of $n$ gives
$$
n= \frac{(e \nu -m \beta+e \varphi)(e \varphi -g \nu -g \varphi - \beta \varphi) \nu }{(e \varphi -g \nu -g \varphi)g(\nu + \varphi)},
$$
while feasibility (\ref{E5_feas}) yields
$$
n \geq \frac{\nu (e \nu -m \beta +e \varphi)}{(\nu + \varphi)g},
$$
which together with the former one implies 
$$
\frac{e \varphi -g \nu -g \varphi - \beta \varphi}{e \varphi -g \nu -g \varphi} \geq 1
$$
from which
$e \varphi -g \nu -g \varphi - \beta \varphi \geq e \varphi -g \nu -g \varphi$ and ultimately
$- \beta \varphi \geq 0$, once again an impossible condition.
The system therefore does not allow limit cycles.

\section{Conclusions}

We have proposed a two-strain model for an obligated mutualistic ecoepidemic system. Although in its generality we were unable to analyse all the
equilibria stability, except by means of numerical simulations, in the two particular cases examined we have shown that no stable equilibria other
than the origin exist. This at first sight appears not to be a good result from the ecological point of view, since is states that in fact
the ecosystem can vanish. One must keep in mind though that as for the underlying demographic model, a saddle in the phase space is present,
given by $E_1$ or its equivalent points. Therefore the system trajectories, as also shown by the simulations, may instead very well grow unbounded.
When the initial conditions do not fall into its basin of attractions, trajectories are repelled away from the saddle
point $E_1$ in case of model (\ref{Mod_1}) as well as from the remaining equilibria $\widetilde E_2$ and $\widetilde E_3$, when we consider model (\ref{Mod_2})
and in both cases tend to grow without bounds.
This is to be ascribed to
the intrinsic limitations of the demographic assumptions of the model, which underlie
the construction of the ecoepidemic model,
or, in other more biological words, to the quadratic type (Holling type I) mutualistic interactions in the ecosystem.
These results should be compared with \cite{EV:07}, where logistic growth is assumed instead of an
exponential mortality, i.e. the symbiosis is not obligated.
In such case the disease does affect the system's behavior, in some cases even favoring the increment of the coexisting populations levels,
at the expense of having part of them infected.

We have also shown that the two strains cannot coexist together, and this result parallels what has been found in \cite{Roman, Elena1, Elena2}.
Further, the diseases do not alter the behavior of the underlying demographic model.
Therefore the disease influence on the obligated mutualistic ecoepidemic systems is clearly less relevant than in ecoepidemic models with other
types of population interactions, in which instead the introduction of the epidemic changes the stability of some of the equilibria,
\cite{EV95,EV01,EV02}. For facultative associations, it is instead known that the results are in line with other ecoepidemic systems of predatory interaction
or competing nature, \cite{NOVA}.
Thus, our future research in two-strained symbiotic systems will aim at removing the obligated mutualism assumption and investigate its consequences.

{\textbf{Acknowledgements}}:
This research was partially supported by
the project ``Metodi numerici in teoria delle popolazioni''
of the Dipartimento di Matematica ``Giuseppe Peano''.


\begin{thebibliography}{11}

\bibitem{ABCDe} G. Allasia, R. Besenghi, R. Cavoretto, A. De Rossi,
Scattered and track data interpolation using an efficient strip searching procedure, Appl. Math. Comput.
{\bf 217} (2011) 5949-5966.

\bibitem{AFKSSK} B. J. Adams, A. Fodor, H. S. Koppenh\"ofer, E. Stackebrandt, S. P. Stock, M. G.,
Klein, Biodiversity and systematics of nematode-bacterium entomopathogens, Biological Control
{\bf 37} (2006) 32-49.

\bibitem{A} J. F. Addicott, Competition in mutualistic systems, in
D.H. Boucher (Editor), The Biology of Mutualism: Ecology and Evolution,
(Croom Helm, London 1985) pp. 217-247.

\bibitem{B} D. H. Boucher, The Biology of Mutualism: Ecology and Evolution,
(Croom Helm, London, 1985).

\bibitem{B-EG} C. Boursaux-Eude, R. Gross, New insights into symbiotic associations between ants
and bacteria, Res. Microbiol. {\bf 151} (2000) 513-519.

\bibitem{C_et_al} E. Caccherano, S. Chatterjee, L. Costa Giani, L. Il Grande, T. Romano, G. Visconti,
E. Venturino, Models of symbiotic associations in food chains, in
Symbiosis: Evolution, Biology and Ecological Effects, Alejandro F. Camis\~ao and Celio C. Pedroso (Editors),
(Nova Science Publishers, Hauppauge, NY, 2012) pp. 189-234.

\bibitem{ICNAAM} R. Cavoretto, S. Chaudhuri, A. De Rossi, E. Menduni,
F. Moretti, M. C. Rodi, E. Venturino,
Approximation of Dynamical System’s Separatrix Curves,
Numerical Analysis and Applied Mathematics ICNAAM 2011, T. Simos, G. Psihoylos, Ch. Tsitouras, Z. Anastassi (Editors),
AIP Conf. Proc. {\bf 1389} (2011) 1220-1223; doi: 10.1063/1.3637836.

\bibitem{Emma} R. Cavoretto, A. De Rossi, E. Perracchione, E. Venturino, Reliable approximation of
separatrix manifolds in competition models with safety niches, to appear in International Journal of Computer
Mathematics 2014, in press. DOI: 10.1080/00207160.2013.867955 

\bibitem{Elena1} E. Elena, M. Grammauro, E. Venturino,
Ecoepidemics with Two Strains: Diseased Prey,
Numerical Analysis and Applied Mathematics ICNAAM 2011, T. Simos, G. Psihoylos, Ch. Tsitouras, Z. Anastassi (Editors),
AIP Conf. Proc. {\bf 1389} (2011) 1228-1231; doi: 10.1063/1.3637838.

\bibitem{Elena2} E. Elena, M. Grammauro, E. Venturino, Predator's alternative food sources do not support ecoepidemics with two-strains-diseased prey,
Network Biology {\bf 3(1)} (2013) 29-44.

\bibitem{FCMCR} L. K. Finkes, A. B. Cady, J. C. Mulroy, K. Clay, J. A. Rudgers,
Plant-fungus mutualism affects spider composition in successional fields,
Ecology Letters {\bf 9} (2006) 347-356.

\bibitem {GL} W. Gan, Z. Lin,
Coexistence and asymptotic periodicity in a competitor-competitor-mutualist model,
J. Math. Anal. Appl. {\bf 337} (2008) 1089-1099.

\bibitem{Gull} F. M. D. Gulland,
The impact of infectious diseases on wild
animal populations - a review, in B.T. Grenfell, A.P. Dobson (Editors)
Ecology of infectious diseases in natural populations, (Cambridge Univ. Press. 1995) pp. 20-51.

\bibitem {GYW} M. Gyllenberg, P. Yan, Y. Wang,
Limit cycles for competitor-competitor-mutualist Lotka-Volterra systems,
Phys. D {\bf 221} (2006) 135-145.

\bibitem{NOVA} M. Haque, E. Venturino, Mathematical models of diseases spreading in symbiotic communities,
in J.D. Harris, P.L. Brown (Editors), Wildlife: Destruction, Conservation and Biodiversity (NOVA Science
Publishers, New York, 2009) pp. 135-179.

\bibitem{HP} J. I. Horiuchi, B. Prithiviraj, H. P. Bais, B. A. Kimball, J. M. Vivanco,
Soil nematodes mediate positive interactions between legume plants and {\it{Rhizobium
bacteria}}, Planta {\bf 222} (2005) 848-857.

\bibitem{J80} D. H. Jantzen, P. De Vries, D. E. Gladstone, M. L. Higgins, T. M. Levinsohn,
Self- and
cross-pollination of Encyclia cordigera (Orchidaceae) in Santa Rosa National Park, Costa Rica,
Biotropica {\bf 12} (1980) 1398-1406.

\bibitem {DDD}  B. W. Kooi, L. D. J. Kuijper, S. A. L. M. Kooijman,
Consequences of symbiosis for food web dynamics, J. Math. Biol. {\bf 49}  (2004) 227-271.

\bibitem {KF02} R. Kumar, H. I. Freedman,
Mathematical analysis in a model of obligate mutualism with food chain population,
Nonlinear Dyn. Syst. Theory {\bf 2} (2002) 25-44.

\bibitem {KF89} R. Kumar, H. I. Freedman,
A mathematical model of facultative mutualism with populations interacting in a food chain,
Math. Biosci. {\bf 97} (1989) 235-261.

\bibitem{MPV} H. Malchow, S. Petrovskii, E. Venturino, Spatiotemporal patterns in Ecology and Epidemiology,
(CRC, 2008).

\bibitem{MK} C. B. Muller, J. Krauss,
Symbiosis between grasses and asexual fungal endophytes, Current Opinion in Plant Biology {\bf 8} (2005) 450-456.

\bibitem{Pa} U. Paszkowski,
Mutualism and parasitism: the yin and yang of plant symbioses,
Current in Plant Biology {\bf 9} (2006) 364-370.

\bibitem {RFA} B.Rai, H. I. Freedman, J. F. Addicott,
Analysis of three-species models of mutualism in predator-prey and competitive systems,
Math. Biosci. {\bf 65} (1983) 13-50.

\bibitem{Roman} F. Roman, F. Rossotto, E. Venturino,
Ecoepidemics with two strains: diseased predators, WSEAS Transactions on Biology and Biomedicine {\bf 8} (2011) 73-85.

\bibitem{SC} F. J. Stewart, C. M. Cavanaugh, Bacterial endosymbioses in {\it{Solemya}} (Mollusca:
Bivalvia)---Model system for studies of symbiont-host adaptation, Antonie van Leeuwenhoek {\bf 90} (2006) 343-360.

\bibitem{EV95} E. Venturino, Epidemics in predator-prey models: disease among the prey, in O.
Arino, D. Axelrod, M. Kimmel, M. Langlais: {\em Mathematical Population
Dynamics: Analysis of Heterogeneity, Vol. one: Theory of Epidemics},
(Wuertz Publishing Ltd, Winnipeg, Canada, 1995) pp. 381-393.

\bibitem{EV01} E. Venturino, The effects of diseases on competing species, {\em Math. Biosc.}
{\bf 174} (2001) 111-131.

\bibitem{EV02} E. Venturino, Epidemics in predator-prey models: disease in the predators,
{\em IMA Journal of Mathematics Applied in Medicine and Biology} {\bf 19} (2002)
185-205.

\bibitem{EV:07} E. Venturino, How diseases affect symbiotic communities, Math. Biosc. {\bf 206} (2007) 11-30.

\bibitem{WF} R. Walker, C. M. J. Ferguson, N. A. Booth, , E. J. Allan,
The symbiosis of {\it{Bacillus subtilis}} L-forms with Chinese cabbage
seedlings inhibits conidial germination of {\it{Botrytis cinerea}},
Letters in Applied Microbiology {\bf 34} (2002) 42-45.

 \bibitem {Z} A. A. S. Zaghrout,
Stability and persistence of facultative mutualism with populations interacting in a food chain, part I,
Appl. Math. Comput. {\bf 45} (1991), 1-15.

\bibitem{ZL} B. G. Zhao, F. Lin, Mutualistic symbiosis between {\it{Bursaphelenchus xylophilus}} and
bacteria of the genus {\it{Pseudomonas}}, For. Path. {\bf 35} (2005) 339-345.

\end{thebibliography}
\end{document}